\documentclass[10pt,a4paper,twoside,fleqn]{ActaStyle}

\def\be{\begin{equation}}
\def\ee{\end{equation}}
\def\bea{\begin{eqnarray}}
\def\eea{\end{eqnarray}}
\def\expon#1{{\rm e}^{#1}}
\def\EqRef#1{(\ref{#1})}
\def\pderop#1{\frac{\partial}{\partial #1}}
\def\Stu{Stuchl\'{\i}k}
\def\d{d}
\def\vec#1{\ifmmode\mathchoice{\mbox{\boldmath$\displaystyle#1$}}
  {\mbox{\boldmath$\textstyle#1$}} {\mbox{\boldmath$\scriptstyle#1$}}
  {\mbox{\boldmath$\scriptscriptstyle#1$}}\else
  \hbox{\boldmath$\textstyle#1$}\fi}

\def\authorheading{Z. Stuchl\'{\i}k}
\def\shorttitle{Static configurations of uniform density
  in spacetimes with $\Lambda\neq0$}

\begin{document}
\pagerange{1}{11}
\title{SPHERICALLY SYMMETRIC STATIC CONFIGURATIONS\\
  OF UNIFORM DENSITY IN SPACETIMES\\
  WITH A NON-ZERO COSMOLOGICAL CONSTANT%
  \footnote{Published in Acta Physica Slovaca \textbf{50}(2) (2000),
    pp.~219--228.}}

\author{Zden\v{e}k Stuchl\'{\i}k\email{Zdenek.Stuchlik@fpf.slu.cz}}%
{Institute of Physics, Faculty of Philosophy and Science,
  Silesian University in~Opava, Bezru\v{c}ovo n\'am.~13,
  CZ-746\,01~Opava, Czech Republic}

\day{}

\abstract{Interior solutions of Einstein's equations with a non-zero
  cosmological constant are given for static and spherically symmetric
  configurations of uniform density. The metric tensor and pressure are
  determined for both positive and negative values of the cosmological
  constant. Limits on the outer radius of the interior solutions are
  established. It is shown that, contrary to the cases of the limits on the
  interior Schwarzschild and Reissner--Nordstr\"om solutions with a zero
  cosmological constant, these limits do not fully coincide with the
  conditions of embeddability of the optical reference geometry associated
  with the exterior (vacuum) Schwarzschild--de~Sitter spacetime.}


\section{Introduction}

Recent observations give strong arguments for the presence of a non-zero
repulsive cosmological constant, $\Lambda>0$. The `concordance' models of
the observations of the fluctuations of the microwave cosmic background and
the measurements of the Hubble constant favor the parabolic model of the
universe, deserved by the inflationary paradigm~\cite{OsSt,KrTu}. The
parabolic model with a zero cosmological constant remains strongly
disfavored in comparison with the parabolic model including a repulsive
cosmological constant, according to an analysis of the new globular cluster
dating and baryon abundance constrains~\cite{Kr}. Moreover, the data of the
high-redshift supernovae indicate a negative deceleration parameter, giving
a direct evidence for a repulsive cosmological constant~\cite{Pe}.

The standard cosmological models with a non-zero cosmological constant were
extensively discussed from the theoretical point of view by
Tolman~\cite{To}, and in connection to observational cosmological
parameters in~\cite{MiThWh}, and recently by B\"orner~\cite{Bo}. Also
properties of the black-hole (and naked-singularity) spacetimes with a
non-zero cosmological constant were discussed extensively (see, e.g.,
\cite{St83,St84,StCa,StBaOsHl}).

Nevertheless, it is interesting to investigate static and spherically
symmetric solutions of Einstein's equations with a non-zero cosmological
constant representing relativistic stars. It was shown by
Schwarzschild~\cite{Sc}, and extensively discussed in~\cite{MiThWh}, that
in the case of a zero cosmological constant the spacetime structure of
relativistic stars with a uniform density can be explicitly given in terms
of elementary functions. The interior Schwarzschild--de~Sitter spacetimes of
uniform density (for a repulsive cosmological constant) were discussed
under some simplifying assumptions in~\cite{To}. Here, it will be shown
that the structure of relativistic stars of uniform density can be
determined by elementary functions for both positive and negative values of
the cosmological constant. It is useful to consider also negative values of
the cosmological constant, because it was recognized recently that the
anti-de~Sitter spacetimes play an important role in the low-energy limit of
the superstring theory~\cite{Sen}. Moreover, it enables us to give a
complete discussion of relations between the restrictions on existence of
the static configurations, and the limits of embeddability of the optical
reference geometry~\cite{AbCaLa}, associated with the vacuum solutions of
Einstein's equations with $\Lambda\neq 0$ describing the spacetime outside
the static configurations.

In Section~\ref{equations}, Einstein's equations with a non-zero
cosmological constant and the conservation law of energy-momentum tensor
are used in the case of spherically symmetric spacetimes to give the
equations of structure of spherically symmetric and static configurations
representing relativistic stars. In Section~\ref{configuration}, the
equations of structure are explicitly integrated for the configurations of
uniform density, and the pressure and metric tensor inside of these
configurations are given. The interior geometry with a non-zero
cosmological constant is always smoothly matched to an appropriately chosen
exterior vacuum Schwarzschild--de~Sitter or Schwarzschild--anti-de~Sitter
geometry with the same cosmological constant.  Reality conditions on the
existence of these static configurations, giving limits on their outer
radius, are determined.

Concluding remarks, concerning relations of the interior
Schwarzschild--de~Sitter spa\-ce\-time presented in this paper to the general
Stephani solutions of Einstein's equations, and relations between the limits
on the outer radius of the static configurations and the reality conditions of
embeddability of optical reference geometry related to the external vacuum
Schwarzschild--de~Sitter (and Schwarzschild--anti-de~Sitter) spacetimes, are
presented in Section~\ref{remarks}.

\section{Equations of structure}  \label{equations}

In terms of the standard Schwarzschild coordinates ($t, r, \theta, \phi$), the
line element of spherically symmetric, static spacetimes can be given in the
form
\be                                                                 \label{1}
  \d s^2 = - \expon{2 \Phi}\d t^2 + \expon{2 \Psi}\d r^2 + r^2
    (\d\theta^2 + \sin^2 \theta\,\d\phi^2)
\ee
with just two unknown functions, $\Phi(r)$ and $\Psi(r)$. (In the following,
we shall use the geometric system of units with $c=G=1$.) To high precision,
the matter inside any star can be assumed to be a perfect fluid with the
stress-energy tensor in the form
\be                                                                 \label{2}
  T^{\mu\nu} = (\rho + p) u^{\mu} u^{\nu} + p g^{\mu\nu}
\ee
where, in rest-frame of the fluid, $\rho = \rho (r)$ is the density of
mass-energy, $p = p(r)$ is the isotropic pressure, $u^{\mu} = u^{\mu}(r)$ is
the 4-velocity of the fluid, and $g^{\mu \nu}$ is the metric tensor of the
spacetime. In a static star, each element of the fluid must remain at rest in
the static coordinate system. Therefore,
\be                                                                 \label{3}
  u^r = \frac{\d r}{\d\tau} = 0\quad  
  u^{\theta}= \frac{\d\theta}{\d\tau} = 0\quad  
  u^{\phi} = \frac{\d\phi}{\d\tau} = 0
\ee
and from the normalization condition it follows that
\be                                                                 \label{4}
  u^t = \frac{\d t}{\d\tau} = \expon{-\Phi}.
\ee
The fluid have to be characterized by an equation of state $p = p(\rho)$, or,
in an implicit form, by the number density in rest-frame of the fluid $n =
n(r)$, and the functional dependence of $\rho$ and $p$ on $n$: $\rho =
\rho(n)$, $p = p(n)$.

The structure of a relativistic star is then determined by Einstein's field
equations, $G_{\mu\nu} \equiv R_{\mu\nu} - Rg_{\mu\nu} + \Lambda g_{\mu\nu} =
8\pi T_{\mu\nu}$, and by the law of local energy-momentum conservation,
$T^{\mu\nu}_{\hphantom{\mu\nu};\nu} = 0$. It is convenient to express the
equations in terms of the components on the orthonormal tetrad of 4-vectors
carried by the fluid elements:
\bea
  & &\vec{e}_{(t)}      = \frac{1}{\expon{\Phi}} \pderop{t}      \label{5} \\
  & &\vec{e}_{(r)}      = \frac{1}{\expon{\Psi}} \pderop{r}      \label{6} \\
  & &\vec{e}_{(\theta)} = \frac{1}{r} \pderop{\theta}            \label{7} \\
  & &\vec{e}_{(\phi)}   = \frac{1}{r\sin\theta} \pderop{\phi}.   \label{8}
\eea
Projection of $T^{\mu\nu}_{\hphantom{\mu\nu};\nu} = 0$ orthogonal to $u^{\mu}$
(by the projection tensor $P^{\mu\nu} = g^{\mu\nu} + u^{\mu} u^{\nu}$) gives
the relevant equation
\be                                                                 \label{9}
  (\rho + p) \frac{\d\Phi}{\d r} = - \frac{\d p}{\d r}
\ee
which is the equation of hydrostatic equilibrium describing the balance
between gravitational force and pressure gradient.

There are two relevant structure equations following from the Einstein
equations. These are determined by the $(t)(t)$ and $(r)(r)$ tetrad components
of the field equations (the $(\theta)(\theta)$ and $(\phi)(\phi)$ components
give dependent equations).

First we shall discuss the $(t)(t)$ component:
\be                                                                \label{10}
  G_{(t)(t)} = \frac{1}{r^2} - \frac{\expon{-2 \Psi}}
  {r^2} - \frac{1}{r}
  \frac{\d}{\d r}\,\expon{-\Psi} - \Lambda = 8 \pi \rho.
\ee
We can transfer it into the form
\be                                                                \label{11}
  \frac{\d}{\d r}\left[r \left(1-\expon{-2 \Psi} \right) - \frac{1}{3}\Lambda r^3
  \right] = \frac{\d}{\d r} 2m (r)
\ee
where
\be                                                                \label{12}
  m(r) = \int^r_0 {4 \pi r^2 \rho \, \d r}.
\ee
The integration constant in \EqRef{12} is chosen to be $m(0) = 0$, because it
means the spacetime geometry smooth at the origin (see~\cite{MiThWh}). Then we
find the relation
\be                                                                \label{13}
  \expon{2 \Psi} = \left[1 - \frac{2m(r)}{r} - \frac{1}{3}\Lambda  r^2
  \right]^{-1}.
\ee

The $(r)(r)$ component of the field equations reads:
\be                                                                \label{14}
  G_{(r)(r)} = - \frac{1}{r^2} + \frac{\expon{-2\Psi}}{r^2} + \frac{2
  \expon{-2 \Psi}}{r^2} \, \frac{\d\Phi}{\d r} + \Lambda = 8 \pi p.
\ee
Using \EqRef{13}, we obtain the relation
\be                                                                \label{15}
  \frac{\d\Phi}{\d r} = \frac{m(r)-\frac{1}{3}\Lambda  r^3 + 4 \pi p r^3}{r
  \left[r-2m(r) - \frac{1}{3}\Lambda  r^3 \right]}
\ee
which enables us to put the equation of hydrostatic equilibrium \EqRef{9} into
the Tolman--Oppenheimer--Volkoff (TOV) form modified by the presence of a
non-zero cosmological constant:
\be                                                                \label{16}
  \frac{\d p}{\d r} = - (\rho + p) \frac{\left[m(r) - \frac{1}{3}\Lambda  r^3 + 4
  \pi p r^3 \right]}{r \left[r-2m(r) - \frac{1}{3}\Lambda  r^3 \right]}.
\ee

For realistic equations of state, the equations of stellar structure can be
integrated only numerically. However, the equations of structure can be
integrated analytically for some idealized and ad hoc equations of state. We
shall consider in the next section one of the most useful analytic solutions.

\section{Relativistic configurations of uniform density}  \label{configuration}

Relativistic model of a star of uniform density
\be
  \rho = {\rm const}\ {\rm for\ all}\ p
\ee
has been analytically integrated and discussed in the case of a zero
cosmological constant~\cite{MiThWh,Sc}. Here, it will be shown that the
equations of structure can be analytically integrated even in the case of a
non-zero cosmological constant. We shall discuss the cases of both repulsive,
$\Lambda > 0$, and attractive, $\Lambda < 0$, cosmological constant.

Recall that in the case of a relativistic star with $\rho = {\rm const}$, it
is not necessary to use the unrealistic notion of an `incompressible fluid'.
One can think of the fluids with pressure growing as radius decreases, having
a composition that varies from one radius to another, and  being
`hand-tailored'~\cite{MiThWh}.

Assuming $\rho = {\rm const}$, we can integrate the structure equations
analytically. First, we obtain from the mass formula \EqRef{12} that
\be                                                                \label{17}
  m(r) = \frac{4 \pi}{3} \rho r^3.
\ee
At the surface of the star ($r = R$), we obtain the total mass of the
star
\be                                                                \label{18}
  M = m(R) = \frac{4 \pi}{3} \rho R^3.
\ee
Now, we can easily find the radial component of the metric tensor:
\be                                                                \label{19}
  \expon{2 \Psi(r)} = \left(1 - \frac{r^2}{\alpha^2} \right)^{-1}
\ee
where we have introduced a new parameter by the relation
\be                                                                \label{20}
  \frac{1}{\alpha^2} = \frac{1}{3}(8 \pi \rho + \Lambda).
\ee
At the surface of the star, there is
\be                                                                \label{21}
  \expon{2 \Psi (R)} = \left(1 - \frac{R^2}{\alpha^2}\right)^{-1}= \left(1 -
  \frac{2M}{R} - \frac{1}{3}\Lambda  R^2 \right)^{-1}
\ee
and we can see immediately that the radial metric coefficient of the interior
spacetime is smoothly matched to the corresponding metric coefficient of the
exterior Schwarzschild--de~Sitter (or Schwarzschild--anti-de~Sitter) spacetime
of the mass parameter $M = m(R)$.

If $\rho = {\rm const}$, the modified TOV equation \EqRef{16} reduces into the
form
\be                                                                \label{22}
  \frac{\d p}{(p+ \rho) \left(3p + \rho - \Lambda/4\pi \right)}=
  - \frac{4 \pi}{3} \, \frac{r\, \d r}{\left(1 - r^2/\alpha^2 \right)}
\ee
which have to be integrated from the surface of the star ($r=R$), where $p(R)
= 0$, down to the center of the star at $r=0$.

For a non-zero cosmological constant we find the pressure at a radius $r$ to
be given by the relation
\be                                                          \label{23}
  p(r) = \frac{\rho \left(\rho - \Lambda/4\pi \right) \left[
  \left(1 - r^2/\alpha^2 \right)^{1/2} - \left(1 -
  R^2/\alpha^2 \right)^{1/2} \right]}
  {3 \rho \left(1 - R^2/\alpha^2 \right)^{1/2} - \left(\rho -
  \Lambda/4\pi \right) \left(1- r^2/\alpha^2 \right)^{1/2}},
\ee
and for $\Lambda = 0$ this relation reduces to the well known formula given, e.g.,
in~\cite{MiThWh}. The maximum pressure is at the center of the star, where
\be                                                                \label{24}
  p_{\rm c} = p(r=0) = \frac{\rho \left(\rho- \Lambda/4\pi \right)
  \left[1 - \left(1- R^2/\alpha^2 \right)^{1/2} \right]}
  {3 \rho \left(1 - R^2/\alpha^2 \right)^{1/2} - \left(\rho -
  \Lambda/4\pi \right)}.
\ee
For fixed energy-density, $\rho$, and cosmological constant, $\Lambda$, the
central pressure increases
monotonously
as the outer radius, $R$, increases, and also the mass, $M$, and the `strength
of gravity', $2M/R = \frac{8}{3}\pi\rho R^2$, increase. (Naturally, as more
matter is added to the star, a greater pressure is required to support it.)

The pressure at any relativistic star must be finite and positive. The
restrictions
\bea
  & &\rho - \frac{\Lambda}{4\pi} \geq 0                       \label{25} \\
  & &3 \rho \left(1- \frac{R^2}{\alpha^2} \right)^{1/2} - \left(\rho -
  \frac{\Lambda}{4 \pi} \right) \geq 0                     \label{26}
\eea
yield limits on the allowed values of outer radii $R$ of the stars. The
equality in \EqRef{26} determines limiting configurations with a divergent
central pressure. Substituting for $\rho$ and $ \alpha^2$ from \EqRef{18}, and
\EqRef{20}, respectively, and introducing new dimensionless cosmological and
radius parameters by the relations
\bea
  y &=& \frac{1}{3}\Lambda  M^2                                 \label{27} \\
  x &=& \frac{R}{M}                                             \label{28}
\eea
the condition \EqRef{26} can be transformed into the relation
\be                                                                \label{29}
  [y-y_+(x)] [y-y_-(x)] \leq 0
\ee
where
\be                                                                \label{30}
  y_{\pm}(x) \equiv \frac{2x-9 \pm 3 \left| 2x-3\right| }{2x^4}.
\ee

For the cosmological repulsion ($y>0$) only the function
\be                                                                \label{31}
  y_+(x) = \frac{4x - 9}{x^4}
\ee
is relevant (at $x \geq \frac{9}{4}$). (If $x= \frac{9}{4}$, we arrive at the
well known limit of the interior Schwarzschild solutions~\cite{MiThWh}.)
However, the validity of the condition \EqRef{29} is restricted to the region
up to the maximum of $y_+(x)$, given by \EqRef{31}. It is located at $x_{\rm
max} = 3$, where $y_{\rm max} = \frac{1}{27}$. At $x \geq x_{\rm max} = 3$,
the relevant condition is \EqRef{25} which determines a critical value of the
cosmological constant for a given mass parameter $M$. In terms of the
dimensionless parameters $x$ and $y$, it implies the condition
\be                                                                \label{32}
  y \leq y_{\rm stat} \equiv \frac{1}{x^3}.
\ee
Notice that for $y = y_{\rm stat}$, the outer radius of the star is located
just at the so called static radius $r_{\rm s}$ of the corresponding external
Schwarzschild--de~Sitter spacetime. At $r=r_{\rm s}$, the gravitational
attraction acting on a test particle is just compensated by the cosmological
repulsion~\cite{St84}. (At $r>r_{\rm s}$, the repulsion prevails, and a static
configuration is possible only with a surface stress acting inwards. We shall
not consider such a situation.)

For an attractive cosmological constant, $\Lambda < 0$, the relations
\EqRef{25} and \EqRef{26} have to be satisfied again, but we obtain an other
family of critical values of the cosmological constant, given by the condition
$1/\alpha^2 = 0$. In terms of the dimensionless parameters $x$ and $y$, we
arrive at
\be                                                                \label{33}
  y_{\rm crit} = - \frac{2}{x^3};
\ee
in terms of the constant density $\rho$, the critical value of the
cosmological constant is given by
\be
  \Lambda_{\rm crit} = -8 \pi \rho.
\ee
Now, we have to distinguish the cases $y > y_{\rm crit}, y < y_{\rm crit}$,
and $y = y_{\rm crit}$.

If $y > y_{\rm crit}$ ($1/\alpha^2 > 0$), the relations \EqRef{29} and
\EqRef{30} are valid. Recall that at $x \geq \frac{3}{2}$, there is $y_-(x) =
y_{\rm crit} (x)$, while at $x \leq \frac{3}{2}$, there is $y_+ (x) = y_{\rm
crit} (x)$. For $x = \frac{3}{2}, y_-(x) = y_+(x) = -\frac{16}{27}$.
Therefore, in addition to $y> y_{\rm crit} \equiv -2/x^3$, there must be
satisfied the condition
\be
  -\frac{2}{x^3}\leq y \leq \frac{4x-9}{x^4}                     \label{add1}
\ee
at $x>3/2$, and
\be
  \frac{4x-9}{x^4} \leq y \leq -\frac{2}{x^3}
\ee
at $x < 3/2$.

If $y<y_{\rm crit}$ ($1/ \alpha^2 <0$), the relation \EqRef{29} has to be
replaced by
\be
  [y-y_+(x)] [y-y_-(x)] \geq 0.
\ee
In addition to $y<y_{\rm crit} \equiv -2/x^3$, we obtain the conditions
\be
  y\leq -\frac{2}{x^3}\quad {\rm or} \quad y\geq \frac{4x-9}{x^4}
\ee
at $x>3/2$, and
\be
  y\geq - \frac{2}{x^3}\quad {\rm or}\quad y\leq \frac{4x-9}{x^4} \label{add2}
\ee
at $x<3/2$.  It follows from the conditions \EqRef{add1}--\EqRef{add2} that
for $\Lambda <0$ the static configurations are allowed at radii satisfying
the condition
\be
  y \leq \frac{4x-9}{x^4}.
\ee

In the special case when the outer radii of the static configurations are
determined by the condition $y=y_{\rm crit} (1/ \alpha^2 = 0)$, the
pressure is given by
\be
    p(r) = \frac{2 \pi \rho^2 \left(R^2-r^2 \right)}{1 - 2 \pi \rho \left(R^2
  - r^2 \right)}.
\ee
In term of the dimensionless parameters, the central pressure of this
special class of solutions is determined by the relation
\be
  p_{\rm c} = p(0) = \frac{3 \rho}{2x-3} = -\frac{3 \Lambda_{\rm crit}}{8\pi
  (2x-3)}.
\ee
Clearly, the special class of static configurations with $y=y_{\rm crit}$
is allowed for $x\geq 3/2$ only. Values of the cosmological parameter $y$
must be restricted by the condition
\be
  -\frac{16}{27} \leq y < 0.
\ee


%

Finally, we determine the time component of the internal metric tensor, using
the boundary condition of smooth matching of the internal solution onto the
external time metric coefficient at $r=R$:
\be                                                                \label{39}
  \expon{2 \Phi (R)} = \left(1 - \frac{2M}{R} -
  \frac{1}{3}\Lambda  R^2 \right).
\ee
The function $\Phi (r)$ can be found from Eq.\,\EqRef{15} by using the relation
for the pressure as a function of radius. If $1/\alpha^2 \neq 0$, we arrive at
the expression
\bea
  \expon{\Phi(r)} &=& \frac{9M}{6M+ \Lambda R^3} \left(1-\frac{2M}{R} -
  \frac{1}{3}\Lambda  R^2 \right)^{1/2} -                       \nonumber  \\
  & & \frac{3M- \Lambda R^3}{6M+ \Lambda
  R^3} \left(1- \frac{2Mr^2}{R^3} - \frac{1}{3}\Lambda  r^2 \right)^{1/2}
                                                                   \label{40}
\eea
which holds at $r \leq R$ equally for both cases $y>y_ {\rm crit}$ and
$y<y_{\rm crit}$. At $r=R$ the relation \EqRef{40} really reduces to
\EqRef{39}. We can convince ourselves that the reality condition
$\expon{\Phi(r)} \geq 0$ will be satisfied under exactly the same
circumstances, as those derived above for the positivity and finiteness of the
pressure function $p(r)$.

In the special case $y=y_{\rm crit}$ ($1/\alpha^2 = 0$), we obtain
\be                                                                \label{41}
  \expon{\Phi(r)} = 1 + \frac{3M}{2R} \left(\frac{r^2}{R^2} - 1 \right).
\ee
Notice that now
\be                                                                \label{42}
  \expon{\Psi(r)} = 1
\ee
for all $r \leq R$, and the space sections $t = {\rm const}$ have purely
3-dimensional Euclidean geometry. We can immediately see that also in this
special situation the reality condition $\expon{\Phi(r)} \geq 0$ yields the
condition $R \geq \frac{3}{2} M$, obtained from the properties of $p(r)$.

\section{Concluding remarks}  \label{remarks}

The solution of Einstein's equations presented above is a particular case
of the solutions found by Stephani (\cite{Ste}) as one of the metrics which
can be embedded in a flat five-dimensional space, or equivalently, the
Krasi\'nski metrics representing spacetimes made of $0(3)$-symmetric
three-dimensional spacelike hypersurfaces strung onto a timelike line
orthogonal to them \cite{Kra}. These solutions were derived under
assumption that the source in the field equations is a perfect fluid.
Following Krasi\'nski~\cite{Kra}, we can write the general geometry in the
form
\bea
  \d s^2 & = & -D^2 (r,t)\,\d t^2 + (1+Kr^2)^{-1}\,\d r^2 + r^2 (\d
  \theta^2 + \sin^2 \theta\,\d\phi^2 ) \nonumber \\
  D(r,t) &\equiv & r (A(t) \sin\theta\cos\phi + B(t) \sin\theta \sin\phi +
  C(t) \cos \theta) + \nonumber \\
  & &E(t) \left(1+ Kr^2 \right)^{1/2} + s,                         \label{a}
\eea
with $s=1$ or $0$. The energy density $\epsilon$ and pressure $p_{\rm tot}$
(both including the contribution coming from the cosmological term) are
then given by the relations
\bea
  & &8 \pi \epsilon = 8 \pi \rho + \Lambda = - 3 K             \label{b} \\
  & &8 \pi p_{\rm tot} = 8 \pi p - \Lambda = 3 K
  \left(1-\frac{2s}{3D(r,t)}\right).                            \label{c}
\eea
Notice that for $A=B=C=s=0$, $E=1$, $3K = \Lambda$, we obtain the de~Sitter
solution. The interior Schwarzschild--de~Sitter or Schwarzschild--anti-de
Sitter solutions, given above, can be obtained from \EqRef{a} by the
following procedure: First we must redefine the time coordinate by
\be
  t_{\rm (K)} = \frac{3M}{R^3}\, \alpha^2 \left(1-\frac{R^2}{\alpha^2}
  \right)^{1/2} t                                              \label{d}
\ee
and then we must adjust the parameters of the Krasi\'nski metric by the
relations
\bea
  & &A(t) = B(t) = C(t) = 0\quad s=1      \nonumber \\
  & &K = - \left(\frac{2M}{R^3} + \frac{\Lambda}{3} \right) =-
  \frac{\Lambda}{3} (8 \pi \rho + \Lambda)\quad E = - \frac{\left(
  \rho - \Lambda/4\pi \right)}{3\rho (1+KR^2)^{1/2}}.
                                                                \label{f}
\eea
With these parameters, the relation \EqRef{c} for the pressure reduces
correctly to the relation \EqRef{24}. Notice that the metric is conformally
flat \cite{Kra}; this fact can be helpful in investigations of its
geodesics structure.

The optical reference geometry, introduced in~\cite{AbCaLa}, is very useful in
attempts to understand the character of spherically symmetric spacetimes.
Geodesics of the optical geometry exhibit some interesting physical
properties~-- they coincide with the possible trajectories of light rays,
massive particles require a speed-independent orthogonal thrust in order to
move along them, and gyroscopes transported along them do not precess with
respect to the direction of motion~\cite{Ab}. Moreover, the optical geometry
appears to be very useful in the analysis of a variety of unusual processes
that take place around compact objects~\cite{AbPr,AbMi,AbMiSt}.

Properties of the optical spaces associated with spherically symmetric
spacetimes can be appropriately represented by embedding diagrams of their
central planes into the 3-dimensional Euclidean space. The embedding allows an
accurate treatment of some non-trivial effects~\cite{KrSoAb}. However, the
embedding is possible for a limited part of the optical space only. It is
interesting that in the case of Schwarzschild spacetimes the limit of
embeddability of the optical geometry (\cite{AbCaLa}), $r = \frac{9}{4}M$,
coincides with the minimum possible radius of a static configuration of matter
of uniform density with fixed mass $M$ (\cite{MiThWh}). Similarly, static
configurations of charged matter with the charge parameter $Q=M$ have the
minimum radius $r=M$ (\cite{deFeYuFa}), which coincides with the limit of
embeddability of the optical space in the case of Reissner--Nordstr\"om
spacetimes~\cite{KrSoAb}.

Kristiansson, Sonego and Abramowicz~\cite{KrSoAb} therefore presented a
conjecture that the minimum radius of embeddability of the optical geometry
into the Euclidean space coincides with the minimum radius of a static
configuration of given spacetime parameters.

It has been shown in~\cite{StHl98a} that the limits of embeddability for the
optical geometry of the Schwarzschild spacetimes with a non-zero cosmological
constant (both positive and negative) are given by the relation
\be                                                                \label{43}
  y = \leq y_{\rm emb} \equiv \frac{4 \tilde{x} - 9}{\tilde{x}^4}
\ee
where $\tilde{x} = r/M$. Clearly, if this condition determines a minimum
radius of embeddability, it coincides with the minimum of corresponding static
configuration of uniform density (see \EqRef{30}), and the conjecture is
confirmed. However, this is not the whole story, because the condition
\EqRef{43} determines also a maximum radius of embeddability for the
spacetimes with a repulsive cosmological parameter $y < \frac{1}{27}$, and
this differs from the maximum radius of the corresponding static
configuration, given by \EqRef{32}.

For the spacetimes with an attractive cosmological constant, $y<0$, there is
no outer limit of embeddability of the optical geometry and the inner limit
coincides with the limit on radius of static configurations of uniform
density. However, there exists a special class ($1/\alpha^2=0$) of the internal
solutions which corresponds to the outer limit of embeddability of the ordinary
induced geometry on $t = {\rm const}$ hypersurfaces \cite{StHl98a}.


A more detailed discussion of the properties of the static configurations with
an uniform density, including the geodesic structure of the internal
spacetimes, will be discussed elsewhere~\cite{StHl98b}. But it is worth to
mention the existence of stable circular photon orbits inside the
configurations having radius $R < 3M$; surprisingly, the limiting radius is
independent of the cosmological constant. The radius $r=3M$, corresponding to
the unstable circular photon orbits in the external Schwarzschild--de~Sitter
and Schwarzschild--anti-de~Sitter spacetimes plays a crucial role in embeddings
of the optical reference geometry associated with these
spacetimes~\cite{StHl98a}.

\section*{Acknowledgements}

This work has been supported by the Czech Republic grants GA\v{C}R
202/96/0206, GA\v{C}R 202/99/0261, and CEZ J10/98/192400004.  The author
would like to thank Prof.\ A. Krasi\'nski for information on the Stephani
solutions.


\begin{thebibliography}{99}

\bibitem{OsSt} Ostriker, J. P., and Steinhart, P. J. (1995). {\it Nature\/}
  {\bf 377}, 600.

\bibitem{KrTu} Krauss, L. M., and Turner, M. S. (1995). {\it Gen.\ Rel.\
  Grav.\/} {\bf 27}, 1137.

\bibitem{Kr} Krauss, L. M. (1998). {\it Ap.\ J.\/} {\bf 501}, 461.

\bibitem{Pe} Perlmutter, S. (1997). {\it Ap.\ J.\/} {\bf 483}, 565.

\bibitem{To} Tolman, R. C. (1969). {\it Relativity, Thermodynamics, and
  Cosmology}, (Oxford: The Clarendon Press).

\bibitem{MiThWh} Misner, C. W., Thorne, K. S., and Wheeler, J. A. (1973).
  {\it Gravitation}, (San Francisco: W H Freeman and Company).

\bibitem{Bo} B\"orner, G. (1993). {\it The Early Universe},
  (Berlin--Heidelberg--New York: Sprin\-ger-Ver\-lag).
  
\bibitem{St83} \Stu, Z. (1983). {\it Bull.\ Astron.\ Inst.\ Czechosl.\/}
  {\bf 34}, 129.
  
\bibitem{St84} \Stu, Z. (1984). {\it Bull.\ Astron.\ Inst.\ Czechosl.\/}
  {\bf 35}, 205.

\bibitem{StCa} \Stu, Z., and Calvani, M. (1991). {\it Gen.\ Rel.\ Grav.\/}
  {\bf 23}, 507.

\bibitem{StBaOsHl} \Stu, Z., Bao, G., \O stgaard, E., and Hled\'{\i}k, S.
  (1998). {\it Phys.\ Rev.\ D\/} {\bf 58}, 084003 (8 pages).
  
\bibitem{Sc} Schwarzschild, K. (1916). {\it Sitzber.\ Deut. Akad.\ Wiss.\ 
    Berlin.\ 
  Kl.\ Math.-Phys.\ Tech.\/} 424. 

\bibitem{Sen} Sen, A. (1998). {\it Developments in superstring theory},
  hep-th/9810356 v 2.
  
\bibitem{AbCaLa} Abramowicz, M. A., Carter, B., and Lasota, J. P. (1988).
  {\it Gen.\ Rel.\ Grav.\/} {\bf 20}, 1173.
  
\bibitem{Ste} Stephani, H. (1967). {\it Commun.\ Math.\ Phys.\/} {\bf 4},
  137.

\bibitem{Kra} Krasi\'nski, A. (1981). {\it Gen.\ Rel.\ Grav.\/} {\bf 13},
  1021.

\bibitem{Ab} Abramowicz, M. A. (1992). {\it Mon.\ Not.\ R. Astron.\ Soc.\/}
  {\bf 256}, 710.
  
\bibitem{AbPr} Abramowicz, M. A., and Prasanna, A. R. (1990). {\it Mon.\ 
    Not.\ R. Astron.\ Soc.\/} {\bf 245}, 720.

\bibitem{AbMi} Abramowicz, M. A., and Miller, J. C. (1990). {\it Mon.\ Not.\
  R. Astron.\ Soc.\/} {\bf 245}, 729.

\bibitem{AbMiSt} Abramowicz, M. A., Miller, J., and \Stu, Z. (1993). {\it
  Phys.\ Rev.\ D\/} {\bf 47}, 1440.

\bibitem{KrSoAb} Kristiansson, S., Sonego, S., and Abramowicz, M. A. (1998).
  {\it Gen.\ Rel.\ Grav.\/} {\bf 30}(2), 275.

\bibitem{deFeYuFa} de Felice, F., Yu, Y., and Fang, J. (1995). {\it Mon.\
  Not.\ R. Astron.\ Soc.\/} {\bf 277}, L17.

\bibitem{StHl98a} \Stu, Z., and Hled\'{\i}k, S. (1999).
  {Phys.\ Rev.\ D} {\bf 60}, 044006 (15~pp.).
  
\bibitem{StHl98b} \Stu, Z., and Hled\'{\i}k, S, and \v{S}olt\'es, J. (1998)
  Geodesics of the interior Schwarzschild--de~Sitter spacetimes, in
  preparation.

\end{thebibliography}
\end{document}